\documentclass[aip,apl,reprint]{revtex4-2}
\usepackage{graphicx}
\usepackage{amsmath}
\usepackage{amssymb}
\usepackage{gensymb}
\usepackage{hyperref}
\usepackage{color}

\begin{document}
	\title{Electrical tuning of robust layered antiferromagnetism in MXene monolayer}
	\author{Xinyu Yang}
	\author{Ning Ding}
	\author{Jun Chen}
	\author{Ziwen Wang}
	\author{Ming An}
	\email{amorn@seu.edu.cn}
	\author{Shuai Dong}
	\email{sdong@seu.edu.cn}
	\affiliation{School of Physics, Southeast University, Nanjing 211189, China}
	\date{\today}
	
\begin{abstract}
A-type antiferromagnetism, with an in-plane ferromagnetic order and the interlayer antiferromagnetic coupling, owns inborn advantages for electrical manipulations but is naturally rare in real materials except in those artificial antiferromagnetic heterostructures. Here, a robust layered antiferromagnetism with a high N\'eel temperature is predicted in a MXene Cr$_2$CCl$_2$ monolayer, which provides an ideal platform as a magnetoelectric field effect transistor. Based on first-principles calculations, we demonstrate that an electric field can induce the band splitting between spin-up and spin-down channels. Although no net magnetization is generated, the inversion symmetry between the lower Cr layer and the upper Cr layer is broken via electronic cloud distortions. Moreover, this electric field can be replaced by a proximate ferroelectric layer for non-volatility. The magneto-optic Kerr effect can be used to detect this magnetoelectricity, even if it is a collinear antiferromagnet with zero magnetization.
\end{abstract}
\maketitle
	
%\textit{Introduction.-}
Two-dimensional (2D) ferroic materials, including both 2D magnetic and polar layers, have become an emerging branch of condensed matter \cite{AM2020apl,TE2015apl,Shuang2021FP,Xiang2019sci}, since the experimental discoveries of ferromagnetism in CrI$_3$, Cr$_2$Ge$_2$Te$_6$, Fe$_3$GeTe$_2$ \cite{huang2017nature,gong2017nature,burch2018nature}, and ferroelectricity in SnTe, CuInP$_2$S$_6$, In$_2$Se$_3$ \cite{Kai2016sci,liu2016NC,cui2018NL}. 
Despite the fast growing number of 2D ferroics as revealed in experiments or predicted in calculations, there remain some tough issues to be solved. For example, due to the reduced coordination numbers in 2D lattices, the magnetic transition temperatures will be relatively low comparing with their three-dimensional (3D) counterparts. Thus most experimentally verified ferromagnetic (FM) Curie temperatures ($T_{\rm C}$'s) are below the room temperature~\cite{Xiang2019sci}. 
	
An alternative solution is to explore 2D antiferromagnets. In general, antiferromagnets are much abundant than ferromagnets, which provide more candidates to improve the transition temperatures. In addition, the importance of antiferromagnetic (AFM) spintronics has been gradually understood in recent years, which can be intrinsically more energy-saving and fast-operating in device applications \cite{jungwirth2016nn,jungwirth2018np}. However, the control and detection of antiferromagnetism remain challenging, since their net magnetization is fully compensated.
	
Different from the plain ferromagnetism, AFM textures can be rather diverse. A very interesting one is the so-called A-type AFM (A-AFM) order, with antiferromagnetically coupled FM layers. This A-AFM order can be operated by electric voltages via the field effect, manifesting a carrier-density driving magnetoelectricity~\cite{dong2013prb}, while other AFM orders are usually inactive to the field effect. For 2D magnets, Huang \textit{et al.} and Wang \textit{et al.} demonstrated the gate control of magnetism and tunneling magnetoresistance in the CrI$_3$ bilayer with the A-AFM order \cite{huang2018NN,wang2018nc}. Similar phenomena have also been reported in other 2D magnetic bilayers \cite{lv2021Nl,ShiJing2018PNAS,tian2021prb}. However, the A-AFM states in all these bilayers rely on the weak van der Waals (vdW) interaction and special stacking configurations, both of which are very subtle and fragile. Natural robust A-AFM materials remain rare, which are urgently needed for nanoscale magnetoelectric devices based on 2D ferroic materials. Furthermore, the mutual exclusion between the tunability and robustness is also a challenging scientific question.
 
MXenes, as an non-vdW family of 2D materials \cite{naguib2011AM,MA2016apl}, may provide an alternative solution to aforementioned questions. With a general chemical formula $M_{n+1}X_nT_x$ ($M$ is an early transition metal, $X$ stands for carbon and/or nitrogen, and $T$ is the surface terminations such as O, OH, F, and Cl), MXenes can naturally accommodate multiple $M$ layers in a unit sheet, even for the thinnest $n$=$1$ case. Also, the neighboring $M$ layers are connected via the $M$-$X$-$M$ chemical bonds, implying much stronger interlayer coupling than aforementioned vdW ones. Thus, the structures of MXenes not only retain the low-dimensional characteristics but also provide a better platform to pursuit robust layered antiferromagnetism. 

Therefore, we will screen candidates from the n = 1 MXenes, which provides $M$-bilayer for possible A-AFM order. However, for many  $M_{2}$NX$_{2}$ candidates, such as Ti$_2$NO$_2$ \cite{kumar2017ACS}, Cr$_2$NO$_2$, Mn$_2$NF$_2$, Mn$_2$N(OH)$_2$ \cite{wang2016JPCC}, they are metallic with ferromagnetic tendency. Luckily, Cr$_2$CCl$_2$ owns desired properties, which has been studied systematically in the present work. Even though its magnetic ground state was already reported to be N\'eel-type antiferromagnetism \cite{li2021JMCC}, the buckled honeycomb Cr lattice can be considered as two stacking triangular layers with the intralayer ferromagnetic coupling and the interlayer AFM coupling, which can mimic the A-type antiferromagnetism. Thus in the following, it is renamed as A$'$-AFM. In fact, similar idea was also used for the [111]-oriented N\'eel-type perovskite \cite{weng2016prl}, which is an effective approach to obtain robust layered antiferromagnetism. Its magnetic transition temperature is estimated to be very high ($\sim1300$ K), but much lower than the previous estimation ($6095$ K) \cite{li2021JMCC}. Despite its robustness, this layered antiferromagnetism can be tuned by external electric field, or a proximate ferroelectric (FE) layer, which gives rise to the non-negligible magneto-optic Kerr effect (MOKE) signals even if its magnetization is fully compensated.

%\textit{Computational methods.-}
First-principles calculations based on the density functional theory (DFT) are performed with the projector augmented-wave (PAW) pseudopotentials as implemented in the Vienna {\it ab initio} Simulation Package (VASP) \cite{kresse1996Prb}. For the exchange-correlation functional, the PBE parametrization of the generalized gradient approximation (GGA) is adopted \cite{perdew1996Prl}, and the Hubbard $U$ is applied using the Dudarev parametrization \cite{dudarev1998Prb}. As reported previously, a correction of $U_{\rm eff}=3$ eV is imposed on Cr's $3d$ orbitals \cite{he2016JMCC}. 

For the monolayer calculation, a vacuum space of $30$ \AA{} thickness is added along the $c$-axis direction to avoid layer interactions. The energy cutoff is fixed to $500$ eV and the $\Gamma$-centered $9\times9\times1$ Monkhorst-Pack \textit{k}-mesh is adopted for the monolayer and heterostructure, which can lead to a well convergence (see Fig.~S1 in the supplementary material ). The convergence criterion for the energy is $10^{-6}$ eV for self-consistent iteration, and the Hellman-Feynman force is set to $0.01$ eV/\AA{} during the structural optimization. The Gaussian smearing (ISMEAR = $0$) in combination with a small SIGMA = $0.05$ is utilized for both structural optimization and self-consistent interactions. Phonopy is adopted to calculate the phonon band structures \cite{togo2015SM}. For the heterostructure calculation, the vdW correction DFT-D2 method is adopted \cite{grimme2006JCC}. 

The $exciting$ code with the time-dependent DFT (TD-DFT) is adopted to calculate the MOKE signal \cite{Gulans2014JPCM, sagmeister2009PCCP}. The spin-orbit coupling (SOC) is considered in the magnetocrystalline anisotropy, dielectric, and MOKE calculations.

The N\'eel temperature $T_{\rm N}$ for Cr$_2$CCl$_2$ is estimated based on the Heisenberg model using DFT-derived spin exchange parameters. A $48 \times 48$ 2D honeycomb lattice with periodic boundary conditions is used in our Markov-chain Monte Carlo (MC) simulations \cite{landau2021guide}. The initial $10^{4}$ MC steps are discarded for thermal equilibrium and the remaining $10^{4}$ MC steps are reserved as statistical average in the simulation. $T_{\rm N}$ is found as a maximum on the temperature dependent specific heat. 

%\textit{Results and discussion.-}
%\subsection{Magnetism of Cr$_2$CCl$_2$ monolayer}
The surfaces of MXenes are usually passivated by anions or chemical ligands. For Cr$_2$CCl$_2$ monolayer with Cl-terminated surfaces, there are four most possible sites for Cl adatoms \cite{khazaei2013AFM}, as shown in Fig.~S2 in the supplementary material. According to our calculations, the model 2 is the most stable structure, which has the lowest energy among all considered configurations  (see Table~S1 in the supplementary material). Therefore, only the model 2 will be studied below in detail. 
	
\begin{figure}
\includegraphics[width=0.48\textwidth]{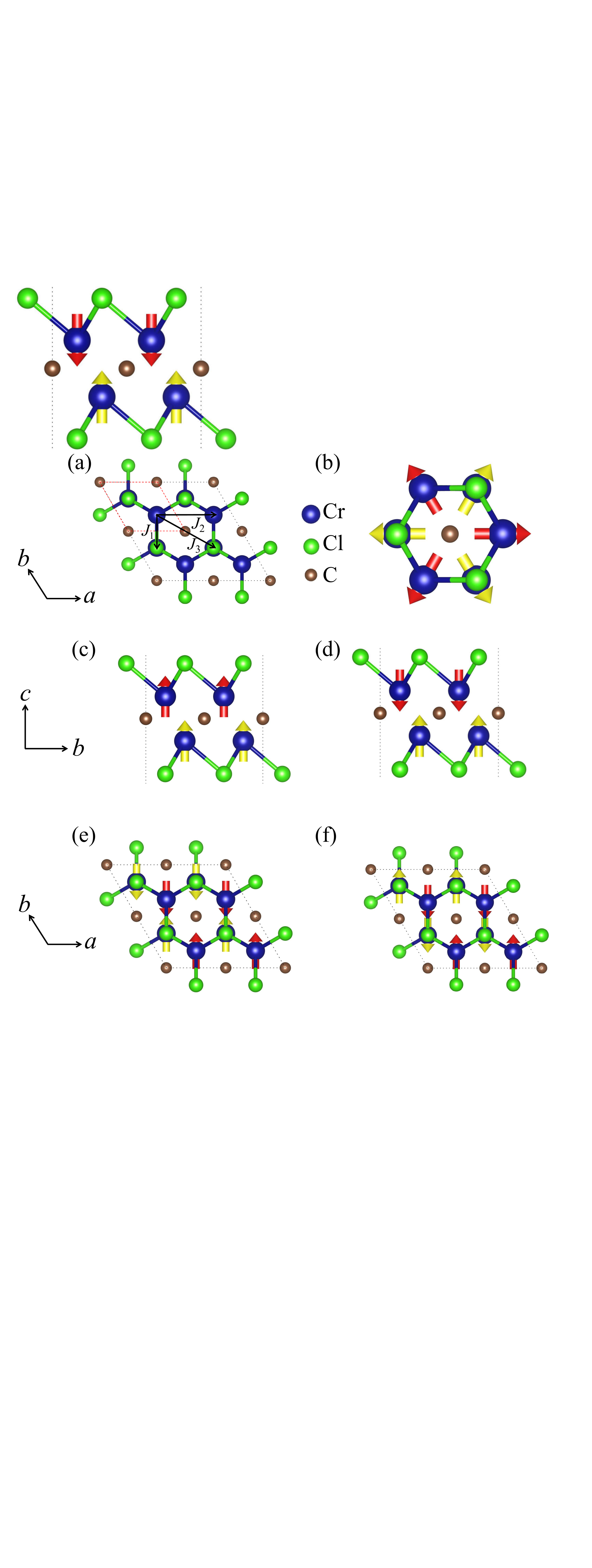}
\caption{(a) The top view of Cr$_2$CCl$_2$ monolayer ($2\times2\times1$ supercell). The primitive cell is indicated by the red dotted rhombus. Exchange $J$'s are also indicated: interlayer $J_1$ and $J_3$, intralayer $J_2$. (b-f) Schematic of five most possible magnetic orders: (b) Y-type noncollinear antiferromagnetism (Y-AFM); (c) Ferromagnetism (FM); (d) A$'$-AFM; (e) Zigzag-type antiferromagnetism (Z-AFM); (f) Stripy-type antiferromagnetism (S-AFM). (c, d) are sideviews and (b, e, f) are topviews. Red and yellow arrows: the spins of upper and lower Cr layers respectively.}
\label{F1}
\end{figure}	

As shown in Fig.~\ref{F1}, in the Cr$_2$CCl$_2$ monolayer, the carbon layer is sandwiched between two chromium layers and the outside surfaces are decorated by chlorines, which possesses a trigonal lattice with the space group (S.G.) $P\bar{3}m1$. Its dynamical stability has been confirmed by phonon calculation, and no imaginary mode appears in the phonon spectrum over the entire Brillouin zone, as shown in Fig.~S3 of the supplementary material.
	
To determine its magnetic ground state, five most possible magnetic orders are compared, as shown in Figs.~\ref{F1}(b-f). According to our calculation, the energy of A$'$-AFM is significantly lower than those of other four (as summarized in Table~\ref{Table 1}), implying the ground state. The optimized in-plane lattice constant is also consistent with previous reported value ($3.269$ \AA{}) \cite{he2016JMCC}.

\begin{table}
	\caption{Five magnetic orders' energies (in units of meV/f.u.) and optimized lattice constants (in units of \AA) of Cr$_2$CCl$_2$ monolayer. The Y-AFM order is calculated in $\sqrt{3}\times\sqrt{3}\times1$ supercell. The FM and A$'$-AFM order use primitive cell, and the other two use $1\times2\times1$ supercell. The energies are in relative to the A$'$-AFM one. The band gap is in units of eV. $M$ is the magnetic moment (in units of $\mu_{\rm B}$).}
	\begin{tabular*}{0.48\textwidth}{@{\extracolsep{\fill}}lcccccc}
		\hline \hline
		Order & Energy & S.G. & $a$ & $b$ & $M$(Cr)& Gap \\
		\hline
		A$'$-AFM & 0     & $P\bar{3}m1$ & 3.258  &   &   $\pm3.099$    & 1.67 \\		
		FM     & 495.5 & $P\bar{3}m1$& 3.270 &   &  3.135  &  0    \\
		Y-AFM  & 547.9 & $P\bar{3}m1$ & 5.620 &   &$\pm2.937$  & 0.96 \\
		Z-AFM  & 355.5 & $C2/m$ & 3.272 & 6.456 &$\pm3.121$ &1.16  \\		
		S-AFM & 612.4 & $C2/m$ & 3.243 & 6.508 & $\pm2.812$&0.55\\
		\hline \hline
	\end{tabular*}
	\label{Table 1}
\end{table}		

Based on the optimized structure of its ground state, the exchange couplings are derived by mapping the DFT energy to the Heisenberg model with normalized spins ($|S|=1$). The nearest-neighbor, next-nearest-neighbor, and next-next-nearest-neighbor exchanges (denoted as $J_1$, $J_2$, and $J_3$, respectively) are estimated as $45.7$, $-37.7$, $45.2$ meV, respectively. The chemical bonding Cr-C-Cr makes the $J_3$ coupling even stronger than $J_2$. Both $J_1$ and $J_3$ prefer AFM interlayer coupling, while $J_2$ prefers intralayer ferromagnetism. Such a configuration of $J$'s is not frustrated, which co-stabilize the layered A$'$-AFM order.  These large $J$'s are originated from the half-filled $t_{\rm 2g}$ orbitals of Cr$^{3+}$ and the strong $p$-$d$ hybridization of Cr-C bonds, both of which are advantage for a strong antiferromagnetic coupling. In addition, our $J_1$/$J_2$ is close to the values reported in the previous study \cite{he2016JMCC}, although they did not consider $J_3$.

\begin{figure}
\includegraphics[width=0.48\textwidth]{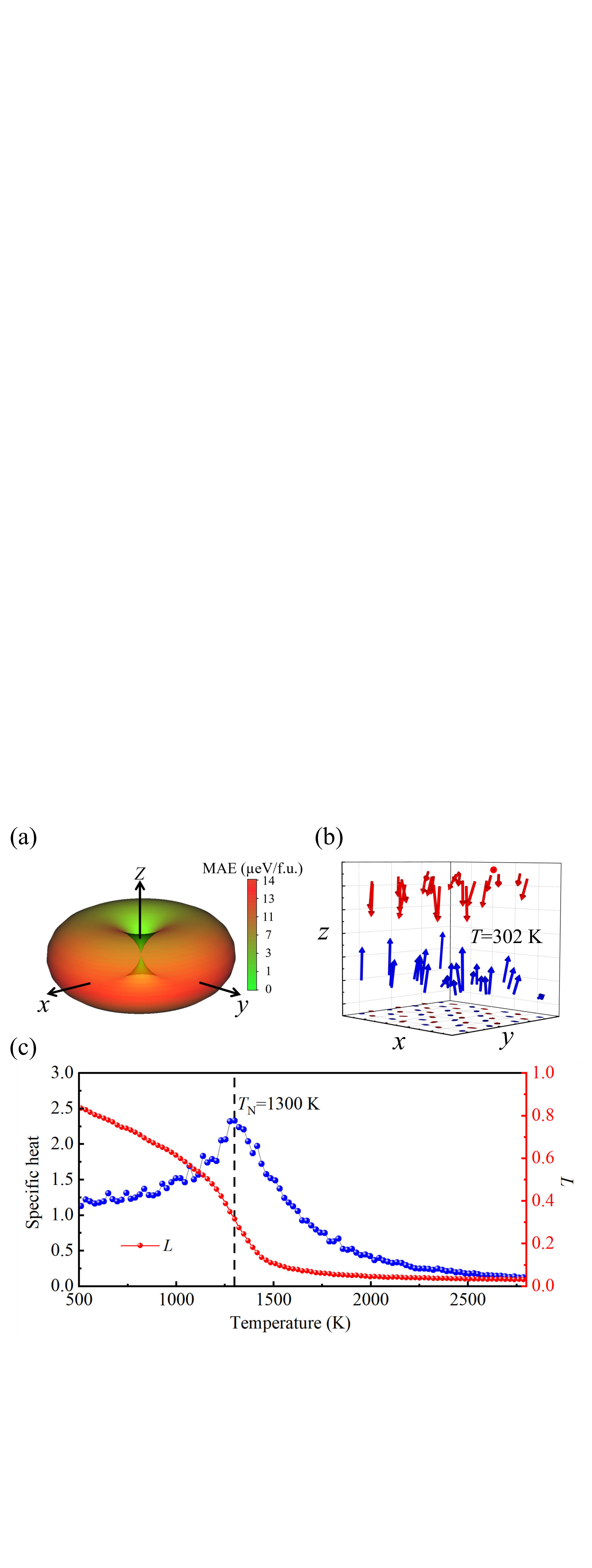}
\caption{(a) The DFT calculated MAE as a function of spin orientation. $z$-axis is along the [001] crystallographic orientation. (b) A typical MC snapshot at $302$ K, showing the A$'$-AFM order with some cantings due to thermal fluctuation. (c) MC simulated specific heat and normalized antiferromagnetic order parameter $L$ as a function of temperature. The antiferromagnetic order parameter is defined as $L$=$S_{u}$-$S_{l}$, where $S$ is the normailzied spin and $u$/$l$ denote the upper and lower layer of Cr bilayers. The peak indicates the transition.}
\label{F2}
\end{figure}

Its magnetic anisotropy is also calculated by rotating the spin orientation. As shown in Fig.~\ref{F2}(a), the magnetic easy axis is along the [001]-axis, i.e. the out-of-plane direction. The magnetocrystalline anisotropy energy (MAE) is estimated to be $14$ $\mu$eV/f.u., due to the weak SOC effect in the Cr$_2$CCl$_2$ monolayer. 

Based on the above DFT-derived coefficients, the MC method was employed to simulate the magnetic transition. A typical MC snapshot at $302$ K, as shown in Fig.~\ref{F2}(b), suggests that an A$'$-AFM order has already been well established at room temperature. In fact, its N\'eel temperature $T_{\rm N}$ is estimated to be $\sim1300$ K, as indicated by the peak of specific heat shown in Fig.~\ref{F2}(c). This high $T_{\rm N}$ is reasonable considering the strong $J$'s, but much lower than the previous estimation ($6095$ K)~\cite{li2021JMCC}, which seems unreasonable.

\begin{figure}
\includegraphics[width=0.48\textwidth]{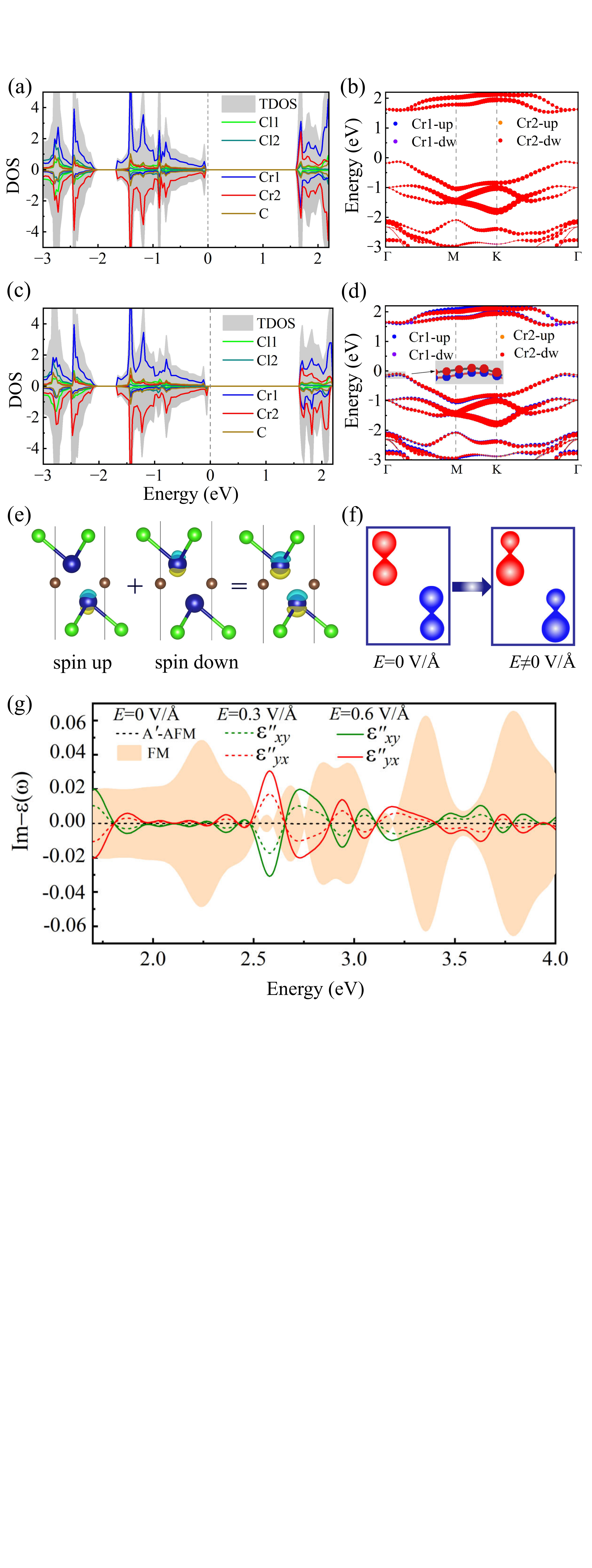}
\caption{(a,b) The DOS and band structure of Cr$_2$CCl$_2$ monolayer. (c,d) The  DOS and band structure of Cr$_2$CCl$_2$ monolayer under a perpendicular electric field ($0.3$ V/\AA). Inset: magnified view of band splitting near the $\Gamma$ point. (e) The charge cloud difference caused by the electric field. Blue: losing electron; yellow: gain of electron. (f) Schematic of electron cloud distortions. (g) Imaginary components of off-digonal dielectric coefficients $\varepsilon''_{xy}$ and $\varepsilon''_{yx}$. Both are zero for the A$'$-AFM phase when there is no electric field. They become nonzero under electric field, and increase with the field. For comparison, the corresponding values of FM phase without field is shown as shadow.}
\label{F3}
\end{figure}

%\subsection{Electric field tuning of A$'$-AFM state}
As mentioned before, such layered antiferromagnetism can be actively coupled with external electric fields along the $c$-axis. The magnetic point group of the Cr$_2$CCl$_2$ monolayer is centrosymmetric $-3'm'$, in which an electric field along the $c$-axis can generate an effective internal magnetic field \cite{zhao2022prl}. This magnetoelectric effect generally works, even here Cr$_2$CCl$_2$ monolayer is a semiconductor with a moderate bandgap [see Figs.~\ref{F3}(a-b)].

After applying an out-of-plane electric field (e.g. $0.3$ V/\AA), the electronic states, i.e. the density of states (DOS) and electronic bands, are slightly split between the spin-up and spin-down channels, as shown in Figs.~\ref{F3}(c-d). This electric field induced splitting is similar to the Zeeman splitting due to the magnetic field. However, this Zeeman-like splitting will not generate a net magnetization at zero temperature, since the existence of bandgap.

This Zeeman-like splitting can also be visualized in real space. As illustrated in Figs.~\ref{F3}(e-f), for both spin-up and spin-down Cr's, the electronic cloud will be distorted by moving against the electric field direction. This distortion breaks the inversion symmetry between the lower layer Cr (spin-up) and upper layer Cr (spin-down), which results in the spin splitting. 

Although no net magnetization is induced by such electronic cloud distortions at zero temperature, this magnetoelectricity can be detected by some sensitive techniques, such as MOKE. In fact, the MOKE has been widely employed as a powerful tool for the characterization of low-dimensional magnetic materials \cite{DM2004apl,Ke2020ACS,huang2017nature,gong2017nature,burch2018nature,huang2018NN}. 

Generally, the signal of MOKE is associated with the off-diagonal components of the optical conductivity tensors $\sigma$ \cite{Ke2020ACS}. The optical conductivity has a relationship with the dielectric tensor $\varepsilon$: $\varepsilon_{ij}(\omega)=\delta_{ij}+i\frac{4\pi}{\omega}\sigma_{ij}(\omega)$~\cite{sangalli2012prb}. Therefore, the presence of MOKE directly depends on the imaginary components of the dielectric function. Under the electric field along the $c$-axis, the original magnetic point group $-3'm'$ is decreased to $3m'$, whose dielectric tensor can be generally expressed as~\cite{born2013principles}:
\begin{equation}
\varepsilon={\left[\begin{array}{ccc}
\varepsilon_{xx}&\varepsilon_{xy}&0\\
-\varepsilon_{xy}&\varepsilon_{xx}&0\\
0&0&\varepsilon_{zz}\\
\end{array}
\right]}.
\label{3}
\end{equation}	
	
Its frequency dependent dielectric function has been calculated. As shown in Fig.~\ref{F3}(g), for the A$'$-AFM phase, the imaginary components of off-diagonal $\varepsilon''_{xy}$ stay zero in the absence of an electric field. In contrast, it is nonzero for the FM phase, implying a detectable MOKE signal. The most interesting result is that the imaginary components of $\varepsilon''_{xy}$ are also nonzero for the A$'$-AFM phase under an external electric field, although the net magnetization remains zero. As expected, larger electric field leads to larger amplitude of imaginary components of $\varepsilon''_{xy}$. In short, the Zeeman-like band splitting in the Cr$_2$CCl$_2$ monolayer induced by an electric field can be detected via magneto-optical activity, even in the absence of net magnetization.

%\subsection{Ferroelectric tuning of A$'$-AFM state}
An alternative and feasible approach is to use a proximate layer to replace the external electric field, which can be even nonvolatile for the magnetoelectric switching. 

To demonstrate this idea, a 2D vdW heterostructure Cr$_2$CCl$_2$/Sc$_2$CO$_2$ is constructed. Here the Sc$_2$CO$_2$ monolayer is a ferroelectric MXene, with a spontaneous out-of-plane polarization of $1.60$ $\mu$C/m$^2$ \cite{Chandrasekaran2017nl}. Also, the in-plane lattice geometry matches well between Sc$_2$CO$_2$ and Cr$_2$CCl$_2$ monolayers: similar trigonal framework and proximate lattice constants ($a = 3.427$ \AA{} for Sc$_2$CO$_2$ and $a = 3.258$ \AA{} for Cr$_2$CCl$_2$ according to our DFT structural optimization).

\begin{figure*}
\includegraphics[width=0.9\textwidth]{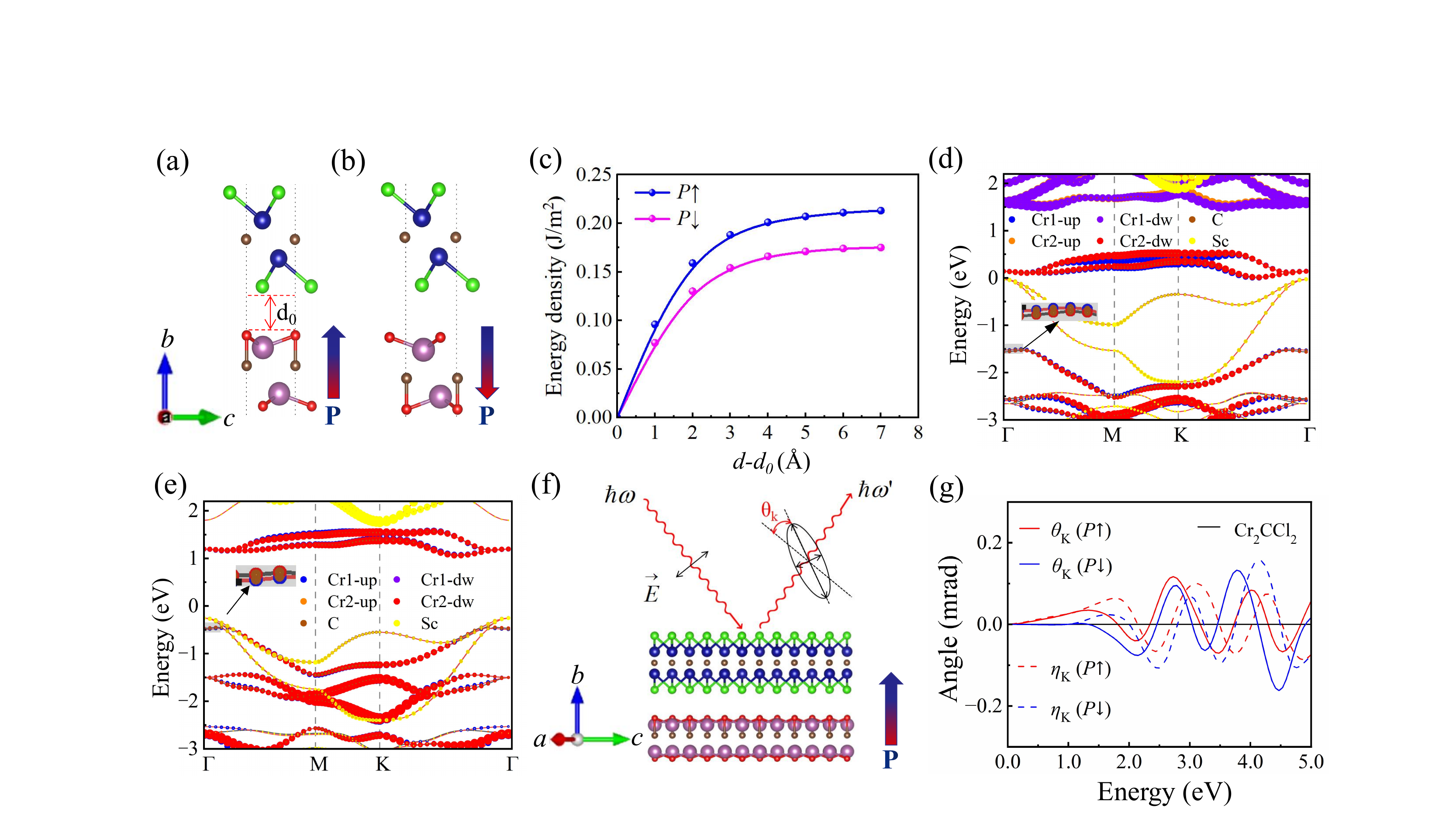}
\caption{(a-b) Side views of Cr$_2$CCl$_2$/Sc$_2$CO$_2$ hetrostructures for the stable configuration A, (a) Cr$_2$CCl$_2$/$P\uparrow$ and (b) Cr$_2$CCl$_2$/$P\downarrow$. (c) The energy dependence of interlayer spacing ($d-d_0$) in the configuration A. (d-e) The calculated band structures of Cr$_2$CCl$_2$/$P\uparrow$ ($P\downarrow$) heterostructures ( the corresponding DOS is shown in Fig.~S5 of the supplementary material). Insets: magnified view of local splittings. The atomic projections are indicated: Cr1/2-up/dw denotes the spin-up/down channel of first/second layer Cr. (f) Schematic illustration of magneto-optical Kerr in Cr$_2$CCl$_2$/$P\uparrow$ heterostructure. Red arrows denote the propagation direction of light, and black double-headed arrows refer to the corresponding linear polarization direction. (g) The calculated MOKE signals of Cr$_2$CCl$_2$/$P\uparrow$ ($P\downarrow$) heterostructures.}
	\label{F4}
\end{figure*}

Considering the FE polarization directions ($P\uparrow$ \textit{vs} $P\downarrow$) and stacking modes (A \textit{vs} B), here four possible structural configurations are considered, as shown in Fig.~S4 of the supplementary material. According to our calculation, the configuration A always has a lower energy than the configuration B, for giving polarization (see Table~S2 in the supplementary material). Hence, only the configuration A will be discussed in the following, as sketched in Figs.~\ref{F4}(a-b). The optimized distances between Cr$_2$CCl$_2$ and Sc$_2$CO$_2$ are $d_0=2.734$ \AA{} for $P\uparrow$ and $d_0=2.868$ \AA{} for $P\downarrow$, respectively. The $P\uparrow$ state is lower in energy than that of $P\downarrow$.

Based on our DFT calculations, the A$'$-AFM phase remains the ground state of Cr$_2$CCl$_2$ upon the substrate polarization (see Tables~S3 in the supplementary material). In addition, the energy dependence of interlayer spacing is shown in Fig.~\ref{F4}(c). The saturation energies of heterostructure are only $0.213$ and $0.175$ J/m$^2$ for $P\uparrow$ and $P\downarrow$ respectively, which are in the range of vdW materials \cite{An2019JPCC}. Thus the coupling between Cr$_2$CCl$_2$ and Sc$_2$CO$_2$ monolayers is the weak vdW interaction instead of the stronger chemical bonding. Furthermore, we have calculated the exchange $J$'s based on the optimized A$'$-AFM structures of Cr$_2$CCl$_2$/$P\uparrow$ and Cr$_2$CCl$_2$/$P\downarrow$ heterostructures, as compared in Table S4 in the supplementary material. It demonstrated that the robust antiferromagnetism and high Néel temperature persist in the heterostructures.

However, a significant tuning of bandgap (from $0.05$ eV for $P\uparrow$ to $1.33$ eV for $P\downarrow$) occurs when switching the polarization of the Sc$_2$CO$_2$ monolayer, as compared in Figs.~\ref{F4}(d-e). This is mostly due to the electrostatic field effect, which largely shifts the conducting band contributed by Cr's empty $3d$ orbitals. As a consequence, the optical properties will be largely different between the $P\uparrow$ and $P\downarrow$ conditions, which will be reflected in the MOKE behavior.

The magnetic point group of the Cr$_2$CCl$_2$/Sc$_2$CO$_2$ heterostructure is $3m'$, which does not change during the polarization switching. The optical conductivity can be expressed as Eq.~S6 in the supplementary material \cite{Ke2020ACS}. The MOKE signal is expected in this heterostructure, characterized by the complex Kerr angle $\phi_{\rm K}$, as sketched in Fig.~\ref{F4}(f). 

The complex Kerr angle $\phi_{\rm K}$ is consisted by the Kerr rotation angle $\theta_{\rm K}$ and Kerr ellipticity $\eta_{\rm K}$: $\phi_{\rm K}=\theta_{\rm K}+i\eta_{\rm K}$. As shown in Fig.~\ref{F4}(g), no MOKE signal appears in pristine Cr$_2$CCl$_2$, but $\phi_{\rm K}$ emerges in the heterostructure. The $\phi_{\rm K}$'s are asymmetric between the $P\uparrow$ and $P\downarrow$ conditions, since these two states are highly asymmetric in this heterostructure. Also, the magnitude of the predicted MOKE signal is within the detectable precision~\cite{kato2004SCI,Lee2016NN}.
	
%\textit{Conclusion.-}
In summary, based on first-principles calculations, we have predicted the electric field induced Zeeman-like splitting of band structures and the magneto-optical Kerr effect in the layered collinear antiferromagnetic Cr$_2$CCl$_2$ monolayer. The effect also occurs in ferroelectric-magnetic heterostructures, such as Cr$_2$CCl$_2$/Sc$_2$CO$_2$. The high magnetic transition temperature of the Cr$_2$CCl$_2$ monolayer makes this magnetoelectric function available at room temperature. Our work opens a promising avenue for future studies of electrical tuning of low-dimensional antiferromagnetic spintronics.

\section*{}
See the supplementary material for more DFT results, including DFT energies, structures, model Hamiltonian, and MOKE equations.

\begin{acknowledgments}
Work was supported by National Natural Science Foundation of China (Grant Nos. 12274069, 12274070, and 11834002). This research work is supported by the Big Data Computing Center of Southeast University.
\end{acknowledgments}

\section*{AUTHOR DECLARATIONS}
\subsection*{Conflict of Interest}
The authors have no conflicts to disclose.

\bibliography{reference}

%apsrev4-2.bst 2019-01-14 (MD) hand-edited version of apsrev4-1.bst
%Control: key (0)
%Control: author (72) initials jnrlst
%Control: editor formatted (1) identically to author
%Control: production of article title (-1) disabled
%Control: page (0) single
%Control: year (1) truncated
%Control: production of eprint (0) enabled
\begin{thebibliography}{43}%
\makeatletter
\providecommand \@ifxundefined [1]{%
 \@ifx{#1\undefined}
}%
\providecommand \@ifnum [1]{%
 \ifnum #1\expandafter \@firstoftwo
 \else \expandafter \@secondoftwo
 \fi
}%
\providecommand \@ifx [1]{%
 \ifx #1\expandafter \@firstoftwo
 \else \expandafter \@secondoftwo
 \fi
}%
\providecommand \natexlab [1]{#1}%
\providecommand \enquote  [1]{``#1''}%
\providecommand \bibnamefont  [1]{#1}%
\providecommand \bibfnamefont [1]{#1}%
\providecommand \citenamefont [1]{#1}%
\providecommand \href@noop [0]{\@secondoftwo}%
\providecommand \href [0]{\begingroup \@sanitize@url \@href}%
\providecommand \@href[1]{\@@startlink{#1}\@@href}%
\providecommand \@@href[1]{\endgroup#1\@@endlink}%
\providecommand \@sanitize@url [0]{\catcode `\\12\catcode `\$12\catcode
  `\&12\catcode `\#12\catcode `\^12\catcode `\_12\catcode `\%12\relax}%
\providecommand \@@startlink[1]{}%
\providecommand \@@endlink[0]{}%
\providecommand \url  [0]{\begingroup\@sanitize@url \@url }%
\providecommand \@url [1]{\endgroup\@href {#1}{\urlprefix }}%
\providecommand \urlprefix  [0]{URL }%
\providecommand \Eprint [0]{\href }%
\providecommand \doibase [0]{https://doi.org/}%
\providecommand \selectlanguage [0]{\@gobble}%
\providecommand \bibinfo  [0]{\@secondoftwo}%
\providecommand \bibfield  [0]{\@secondoftwo}%
\providecommand \translation [1]{[#1]}%
\providecommand \BibitemOpen [0]{}%
\providecommand \bibitemStop [0]{}%
\providecommand \bibitemNoStop [0]{.\EOS\space}%
\providecommand \EOS [0]{\spacefactor3000\relax}%
\providecommand \BibitemShut  [1]{\csname bibitem#1\endcsname}%
\let\auto@bib@innerbib\@empty
%</preamble>
\bibitem [{\citenamefont {An}\ and\ \citenamefont {Dong}(2020)}]{AM2020apl}%
  \BibitemOpen
  \bibfield  {author} {\bibinfo {author} {\bibfnamefont {M.}~\bibnamefont
  {An}}\ and\ \bibinfo {author} {\bibfnamefont {S.}~\bibnamefont {Dong}},\
  }\href@noop {} {\bibfield  {journal} {\bibinfo  {journal} {APL Mater.}\
  }\textbf {\bibinfo {volume} {8}},\ \bibinfo {pages} {110704} (\bibinfo {year}
  {2020})}\BibitemShut {NoStop}%
\bibitem [{\citenamefont {Torun}\ \emph {et~al.}(2015)\citenamefont {Torun},
  \citenamefont {Sahin}, \citenamefont {Singh},\ and\ \citenamefont
  {Peeters}}]{TE2015apl}%
  \BibitemOpen
  \bibfield  {author} {\bibinfo {author} {\bibfnamefont {E.}~\bibnamefont
  {Torun}}, \bibinfo {author} {\bibfnamefont {H.}~\bibnamefont {Sahin}},
  \bibinfo {author} {\bibfnamefont {S.~K.}\ \bibnamefont {Singh}},\ and\
  \bibinfo {author} {\bibfnamefont {F.~M.}\ \bibnamefont {Peeters}},\
  }\href@noop {} {\bibfield  {journal} {\bibinfo  {journal} {Appl. Phys.
  Lett.}\ }\textbf {\bibinfo {volume} {106}},\ \bibinfo {pages} {192404}
  (\bibinfo {year} {2015})}\BibitemShut {NoStop}%
\bibitem [{\citenamefont {Zhou}\ \emph {et~al.}(2021)\citenamefont {Zhou},
  \citenamefont {You}, \citenamefont {Zhou}, \citenamefont {Pu}, \citenamefont
  {Gui},\ and\ \citenamefont {Wang}}]{Shuang2021FP}%
  \BibitemOpen
  \bibfield  {author} {\bibinfo {author} {\bibfnamefont {S.}~\bibnamefont
  {Zhou}}, \bibinfo {author} {\bibfnamefont {L.}~\bibnamefont {You}}, \bibinfo
  {author} {\bibfnamefont {H.~L.}\ \bibnamefont {Zhou}}, \bibinfo {author}
  {\bibfnamefont {Y.}~\bibnamefont {Pu}}, \bibinfo {author} {\bibfnamefont
  {Z.~G.}\ \bibnamefont {Gui}},\ and\ \bibinfo {author} {\bibfnamefont {J.~L.}\
  \bibnamefont {Wang}},\ }\href@noop {} {\bibfield  {journal} {\bibinfo
  {journal} {Front. Phys.}\ }\textbf {\bibinfo {volume} {16}},\ \bibinfo
  {pages} {13301} (\bibinfo {year} {2021})}\BibitemShut {NoStop}%
\bibitem [{\citenamefont {Gong}\ and\ \citenamefont
  {Zhang}(2019)}]{Xiang2019sci}%
  \BibitemOpen
  \bibfield  {author} {\bibinfo {author} {\bibfnamefont {C.}~\bibnamefont
  {Gong}}\ and\ \bibinfo {author} {\bibfnamefont {X.}~\bibnamefont {Zhang}},\
  }\href@noop {} {\bibfield  {journal} {\bibinfo  {journal} {Science}\ }\textbf
  {\bibinfo {volume} {363}},\ \bibinfo {pages} {eaav4450} (\bibinfo {year}
  {2019})}\BibitemShut {NoStop}%
\bibitem [{\citenamefont {Huang}\ \emph {et~al.}(2017)\citenamefont {Huang},
  \citenamefont {Clark}, \citenamefont {Navarro-Moratalla}, \citenamefont
  {Klein}, \citenamefont {Cheng}, \citenamefont {Seyler}, \citenamefont
  {Zhong}, \citenamefont {Schmidgall}, \citenamefont {McGuire}, \citenamefont
  {Cobden} \emph {et~al.}}]{huang2017nature}%
  \BibitemOpen
  \bibfield  {author} {\bibinfo {author} {\bibfnamefont {B.}~\bibnamefont
  {Huang}}, \bibinfo {author} {\bibfnamefont {G.}~\bibnamefont {Clark}},
  \bibinfo {author} {\bibfnamefont {E.}~\bibnamefont {Navarro-Moratalla}},
  \bibinfo {author} {\bibfnamefont {D.~R.}\ \bibnamefont {Klein}}, \bibinfo
  {author} {\bibfnamefont {R.}~\bibnamefont {Cheng}}, \bibinfo {author}
  {\bibfnamefont {K.~L.}\ \bibnamefont {Seyler}}, \bibinfo {author}
  {\bibfnamefont {D.}~\bibnamefont {Zhong}}, \bibinfo {author} {\bibfnamefont
  {E.}~\bibnamefont {Schmidgall}}, \bibinfo {author} {\bibfnamefont {M.~A.}\
  \bibnamefont {McGuire}}, \bibinfo {author} {\bibfnamefont {D.~H.}\
  \bibnamefont {Cobden}}, \emph {et~al.},\ }\href@noop {} {\bibfield  {journal}
  {\bibinfo  {journal} {Nature}\ }\textbf {\bibinfo {volume} {546}},\ \bibinfo
  {pages} {270} (\bibinfo {year} {2017})}\BibitemShut {NoStop}%
\bibitem [{\citenamefont {Gong}\ \emph {et~al.}(2017)\citenamefont {Gong},
  \citenamefont {Li}, \citenamefont {Li}, \citenamefont {Ji}, \citenamefont
  {Stern}, \citenamefont {Xia}, \citenamefont {Cao}, \citenamefont {Bao},
  \citenamefont {Wang}, \citenamefont {Wang} \emph {et~al.}}]{gong2017nature}%
  \BibitemOpen
  \bibfield  {author} {\bibinfo {author} {\bibfnamefont {C.}~\bibnamefont
  {Gong}}, \bibinfo {author} {\bibfnamefont {L.}~\bibnamefont {Li}}, \bibinfo
  {author} {\bibfnamefont {Z.}~\bibnamefont {Li}}, \bibinfo {author}
  {\bibfnamefont {H.}~\bibnamefont {Ji}}, \bibinfo {author} {\bibfnamefont
  {A.}~\bibnamefont {Stern}}, \bibinfo {author} {\bibfnamefont
  {Y.}~\bibnamefont {Xia}}, \bibinfo {author} {\bibfnamefont {T.}~\bibnamefont
  {Cao}}, \bibinfo {author} {\bibfnamefont {W.}~\bibnamefont {Bao}}, \bibinfo
  {author} {\bibfnamefont {C.}~\bibnamefont {Wang}}, \bibinfo {author}
  {\bibfnamefont {Y.}~\bibnamefont {Wang}}, \emph {et~al.},\ }\href@noop {}
  {\bibfield  {journal} {\bibinfo  {journal} {Nature}\ }\textbf {\bibinfo
  {volume} {546}},\ \bibinfo {pages} {265} (\bibinfo {year}
  {2017})}\BibitemShut {NoStop}%
\bibitem [{\citenamefont {Burch}\ \emph {et~al.}(2018)\citenamefont {Burch},
  \citenamefont {Mandrus},\ and\ \citenamefont {Park}}]{burch2018nature}%
  \BibitemOpen
  \bibfield  {author} {\bibinfo {author} {\bibfnamefont {K.~S.}\ \bibnamefont
  {Burch}}, \bibinfo {author} {\bibfnamefont {D.}~\bibnamefont {Mandrus}},\
  and\ \bibinfo {author} {\bibfnamefont {J.-G.}\ \bibnamefont {Park}},\
  }\href@noop {} {\bibfield  {journal} {\bibinfo  {journal} {Nature}\ }\textbf
  {\bibinfo {volume} {563}},\ \bibinfo {pages} {47} (\bibinfo {year}
  {2018})}\BibitemShut {NoStop}%
\bibitem [{\citenamefont {Chang}\ \emph {et~al.}(2016)\citenamefont {Chang},
  \citenamefont {Liu}, \citenamefont {Lin}, \citenamefont {Wang}, \citenamefont
  {Zhao}, \citenamefont {Zhang}, \citenamefont {Jin}, \citenamefont {Zhong},
  \citenamefont {Hu}, \citenamefont {Duan}, \citenamefont {Zhang},
  \citenamefont {Fu}, \citenamefont {Xue}, \citenamefont {Chen},\ and\
  \citenamefont {Ji}}]{Kai2016sci}%
  \BibitemOpen
  \bibfield  {author} {\bibinfo {author} {\bibfnamefont {K.}~\bibnamefont
  {Chang}}, \bibinfo {author} {\bibfnamefont {J.~W.}\ \bibnamefont {Liu}},
  \bibinfo {author} {\bibfnamefont {H.~C.}\ \bibnamefont {Lin}}, \bibinfo
  {author} {\bibfnamefont {N.}~\bibnamefont {Wang}}, \bibinfo {author}
  {\bibfnamefont {K.}~\bibnamefont {Zhao}}, \bibinfo {author} {\bibfnamefont
  {A.~M.}\ \bibnamefont {Zhang}}, \bibinfo {author} {\bibfnamefont
  {F.}~\bibnamefont {Jin}}, \bibinfo {author} {\bibfnamefont {Y.}~\bibnamefont
  {Zhong}}, \bibinfo {author} {\bibfnamefont {X.~P.}\ \bibnamefont {Hu}},
  \bibinfo {author} {\bibfnamefont {W.~H.}\ \bibnamefont {Duan}}, \bibinfo
  {author} {\bibfnamefont {Q.~M.}\ \bibnamefont {Zhang}}, \bibinfo {author}
  {\bibfnamefont {L.}~\bibnamefont {Fu}}, \bibinfo {author} {\bibfnamefont
  {Q.~K.}\ \bibnamefont {Xue}}, \bibinfo {author} {\bibfnamefont
  {X.}~\bibnamefont {Chen}},\ and\ \bibinfo {author} {\bibfnamefont {S.~H.}\
  \bibnamefont {Ji}},\ }\href@noop {} {\bibfield  {journal} {\bibinfo
  {journal} {Science}\ }\textbf {\bibinfo {volume} {353}},\ \bibinfo {pages}
  {274} (\bibinfo {year} {2016})}\BibitemShut {NoStop}%
\bibitem [{\citenamefont {Liu}\ \emph {et~al.}(2016)\citenamefont {Liu},
  \citenamefont {You}, \citenamefont {Seyler}, \citenamefont {Li},
  \citenamefont {Yu}, \citenamefont {Lin}, \citenamefont {Wang}, \citenamefont
  {Zhou}, \citenamefont {Wang}, \citenamefont {He} \emph {et~al.}}]{liu2016NC}%
  \BibitemOpen
  \bibfield  {author} {\bibinfo {author} {\bibfnamefont {F.}~\bibnamefont
  {Liu}}, \bibinfo {author} {\bibfnamefont {L.}~\bibnamefont {You}}, \bibinfo
  {author} {\bibfnamefont {K.~L.}\ \bibnamefont {Seyler}}, \bibinfo {author}
  {\bibfnamefont {X.~B.}\ \bibnamefont {Li}}, \bibinfo {author} {\bibfnamefont
  {P.}~\bibnamefont {Yu}}, \bibinfo {author} {\bibfnamefont {J.~H.}\
  \bibnamefont {Lin}}, \bibinfo {author} {\bibfnamefont {X.~W.}\ \bibnamefont
  {Wang}}, \bibinfo {author} {\bibfnamefont {J.~D.}\ \bibnamefont {Zhou}},
  \bibinfo {author} {\bibfnamefont {H.}~\bibnamefont {Wang}}, \bibinfo {author}
  {\bibfnamefont {H.~Y.}\ \bibnamefont {He}}, \emph {et~al.},\ }\href@noop {}
  {\bibfield  {journal} {\bibinfo  {journal} {Nat. Commun.}\ }\textbf {\bibinfo
  {volume} {7}},\ \bibinfo {pages} {12357} (\bibinfo {year}
  {2016})}\BibitemShut {NoStop}%
\bibitem [{\citenamefont {Cui}\ \emph {et~al.}(2018)\citenamefont {Cui},
  \citenamefont {Hu}, \citenamefont {Yan}, \citenamefont {Addiego},
  \citenamefont {Gao}, \citenamefont {Wang}, \citenamefont {Wang},
  \citenamefont {Li}, \citenamefont {Cheng}, \citenamefont {Li} \emph
  {et~al.}}]{cui2018NL}%
  \BibitemOpen
  \bibfield  {author} {\bibinfo {author} {\bibfnamefont {C.~J.}\ \bibnamefont
  {Cui}}, \bibinfo {author} {\bibfnamefont {W.~J.}\ \bibnamefont {Hu}},
  \bibinfo {author} {\bibfnamefont {X.~X.}\ \bibnamefont {Yan}}, \bibinfo
  {author} {\bibfnamefont {C.}~\bibnamefont {Addiego}}, \bibinfo {author}
  {\bibfnamefont {W.~P.}\ \bibnamefont {Gao}}, \bibinfo {author} {\bibfnamefont
  {Y.}~\bibnamefont {Wang}}, \bibinfo {author} {\bibfnamefont {Z.}~\bibnamefont
  {Wang}}, \bibinfo {author} {\bibfnamefont {L.~Z.}\ \bibnamefont {Li}},
  \bibinfo {author} {\bibfnamefont {Y.~C.}\ \bibnamefont {Cheng}}, \bibinfo
  {author} {\bibfnamefont {P.}~\bibnamefont {Li}}, \emph {et~al.},\ }\href@noop
  {} {\bibfield  {journal} {\bibinfo  {journal} {Nano Lett.}\ }\textbf
  {\bibinfo {volume} {18}},\ \bibinfo {pages} {1253} (\bibinfo {year}
  {2018})}\BibitemShut {NoStop}%
\bibitem [{\citenamefont {Jungwirth}\ \emph {et~al.}(2016)\citenamefont
  {Jungwirth}, \citenamefont {Marti}, \citenamefont {Wadley},\ and\
  \citenamefont {Wunderlich}}]{jungwirth2016nn}%
  \BibitemOpen
  \bibfield  {author} {\bibinfo {author} {\bibfnamefont {T.}~\bibnamefont
  {Jungwirth}}, \bibinfo {author} {\bibfnamefont {X.}~\bibnamefont {Marti}},
  \bibinfo {author} {\bibfnamefont {P.}~\bibnamefont {Wadley}},\ and\ \bibinfo
  {author} {\bibfnamefont {J.}~\bibnamefont {Wunderlich}},\ }\href@noop {}
  {\bibfield  {journal} {\bibinfo  {journal} {Nat. Nanotechnol.}\ }\textbf
  {\bibinfo {volume} {11}},\ \bibinfo {pages} {231} (\bibinfo {year}
  {2016})}\BibitemShut {NoStop}%
\bibitem [{\citenamefont {Jungwirth}\ \emph {et~al.}(2018)\citenamefont
  {Jungwirth}, \citenamefont {Sinova}, \citenamefont {Manchon}, \citenamefont
  {Marti}, \citenamefont {Wunderlich},\ and\ \citenamefont
  {Felser}}]{jungwirth2018np}%
  \BibitemOpen
  \bibfield  {author} {\bibinfo {author} {\bibfnamefont {T.}~\bibnamefont
  {Jungwirth}}, \bibinfo {author} {\bibfnamefont {J.}~\bibnamefont {Sinova}},
  \bibinfo {author} {\bibfnamefont {A.}~\bibnamefont {Manchon}}, \bibinfo
  {author} {\bibfnamefont {X.}~\bibnamefont {Marti}}, \bibinfo {author}
  {\bibfnamefont {J.}~\bibnamefont {Wunderlich}},\ and\ \bibinfo {author}
  {\bibfnamefont {C.}~\bibnamefont {Felser}},\ }\href@noop {} {\bibfield
  {journal} {\bibinfo  {journal} {Nat. Phys.}\ }\textbf {\bibinfo {volume}
  {14}},\ \bibinfo {pages} {200} (\bibinfo {year} {2018})}\BibitemShut
  {NoStop}%
\bibitem [{\citenamefont {Dong}\ and\ \citenamefont
  {Dagotto}(2013)}]{dong2013prb}%
  \BibitemOpen
  \bibfield  {author} {\bibinfo {author} {\bibfnamefont {S.}~\bibnamefont
  {Dong}}\ and\ \bibinfo {author} {\bibfnamefont {E.}~\bibnamefont {Dagotto}},\
  }\href@noop {} {\bibfield  {journal} {\bibinfo  {journal} {Phys. Rev. B}\
  }\textbf {\bibinfo {volume} {88}},\ \bibinfo {pages} {140404(R)} (\bibinfo
  {year} {2013})}\BibitemShut {NoStop}%
\bibitem [{\citenamefont {Huang}\ \emph {et~al.}(2018)\citenamefont {Huang},
  \citenamefont {Clark}, \citenamefont {Klein}, \citenamefont {David},
  \citenamefont {Navarro-Moratalla}, \citenamefont {Seyler}, \citenamefont
  {Wilson}, \citenamefont {McGuire}, \citenamefont {Cobden}, \citenamefont
  {Xiao} \emph {et~al.}}]{huang2018NN}%
  \BibitemOpen
  \bibfield  {author} {\bibinfo {author} {\bibfnamefont {B.}~\bibnamefont
  {Huang}}, \bibinfo {author} {\bibfnamefont {G.}~\bibnamefont {Clark}},
  \bibinfo {author} {\bibfnamefont {D.~R.}\ \bibnamefont {Klein}}, \bibinfo
  {author} {\bibfnamefont {M.}~\bibnamefont {David}}, \bibinfo {author}
  {\bibfnamefont {E.}~\bibnamefont {Navarro-Moratalla}}, \bibinfo {author}
  {\bibfnamefont {K.~L.}\ \bibnamefont {Seyler}}, \bibinfo {author}
  {\bibfnamefont {N.}~\bibnamefont {Wilson}}, \bibinfo {author} {\bibfnamefont
  {M.~A.}\ \bibnamefont {McGuire}}, \bibinfo {author} {\bibfnamefont {D.~H.}\
  \bibnamefont {Cobden}}, \bibinfo {author} {\bibfnamefont {D.}~\bibnamefont
  {Xiao}}, \emph {et~al.},\ }\href@noop {} {\bibfield  {journal} {\bibinfo
  {journal} {Nat. Nanotechnol.}\ }\textbf {\bibinfo {volume} {13}},\ \bibinfo
  {pages} {544} (\bibinfo {year} {2018})}\BibitemShut {NoStop}%
\bibitem [{\citenamefont {Wang}\ \emph {et~al.}(2018)\citenamefont {Wang},
  \citenamefont {Guti{\'e}rrez-Lezama}, \citenamefont {Ubrig}, \citenamefont
  {Kroner}, \citenamefont {Gibertini}, \citenamefont {Taniguchi}, \citenamefont
  {Watanabe}, \citenamefont {Imamo{\u{g}}lu}, \citenamefont {Giannini},\ and\
  \citenamefont {FMorpurgo}}]{wang2018nc}%
  \BibitemOpen
  \bibfield  {author} {\bibinfo {author} {\bibfnamefont {Z.}~\bibnamefont
  {Wang}}, \bibinfo {author} {\bibfnamefont {I.}~\bibnamefont
  {Guti{\'e}rrez-Lezama}}, \bibinfo {author} {\bibfnamefont {N.}~\bibnamefont
  {Ubrig}}, \bibinfo {author} {\bibfnamefont {M.}~\bibnamefont {Kroner}},
  \bibinfo {author} {\bibfnamefont {M.}~\bibnamefont {Gibertini}}, \bibinfo
  {author} {\bibfnamefont {T.}~\bibnamefont {Taniguchi}}, \bibinfo {author}
  {\bibfnamefont {K.}~\bibnamefont {Watanabe}}, \bibinfo {author}
  {\bibfnamefont {A.}~\bibnamefont {Imamo{\u{g}}lu}}, \bibinfo {author}
  {\bibfnamefont {E.}~\bibnamefont {Giannini}},\ and\ \bibinfo {author}
  {\bibfnamefont {A.}~\bibnamefont {FMorpurgo}},\ }\href@noop {} {\bibfield
  {journal} {\bibinfo  {journal} {Nat. Commun.}\ }\textbf {\bibinfo {volume}
  {9}},\ \bibinfo {pages} {2516} (\bibinfo {year} {2018})}\BibitemShut
  {NoStop}%
\bibitem [{\citenamefont {Lv}\ \emph {et~al.}(2021)\citenamefont {Lv},
  \citenamefont {Niu}, \citenamefont {Wu},\ and\ \citenamefont
  {Yang}}]{lv2021Nl}%
  \BibitemOpen
  \bibfield  {author} {\bibinfo {author} {\bibfnamefont {H.~F.}\ \bibnamefont
  {Lv}}, \bibinfo {author} {\bibfnamefont {Y.~J.}\ \bibnamefont {Niu}},
  \bibinfo {author} {\bibfnamefont {X.~J.}\ \bibnamefont {Wu}},\ and\ \bibinfo
  {author} {\bibfnamefont {J.~L.}\ \bibnamefont {Yang}},\ }\href@noop {}
  {\bibfield  {journal} {\bibinfo  {journal} {Nano Lett.}\ }\textbf {\bibinfo
  {volume} {21}},\ \bibinfo {pages} {7050} (\bibinfo {year}
  {2021})}\BibitemShut {NoStop}%
\bibitem [{\citenamefont {Gong}\ \emph {et~al.}(2018)\citenamefont {Gong},
  \citenamefont {Gong}, \citenamefont {Sun}, \citenamefont {Tong},
  \citenamefont {Duan}, \citenamefont {Chu},\ and\ \citenamefont
  {Zhang}}]{ShiJing2018PNAS}%
  \BibitemOpen
  \bibfield  {author} {\bibinfo {author} {\bibfnamefont {S.~J.}\ \bibnamefont
  {Gong}}, \bibinfo {author} {\bibfnamefont {C.}~\bibnamefont {Gong}}, \bibinfo
  {author} {\bibfnamefont {Y.~Y.}\ \bibnamefont {Sun}}, \bibinfo {author}
  {\bibfnamefont {W.~Y.}\ \bibnamefont {Tong}}, \bibinfo {author}
  {\bibfnamefont {C.~G.}\ \bibnamefont {Duan}}, \bibinfo {author}
  {\bibfnamefont {J.~H.}\ \bibnamefont {Chu}},\ and\ \bibinfo {author}
  {\bibfnamefont {X.}~\bibnamefont {Zhang}},\ }\href@noop {} {\bibfield
  {journal} {\bibinfo  {journal} {Proc. Natl. Acad. Sci.}\ }\textbf {\bibinfo
  {volume} {115}},\ \bibinfo {pages} {8511} (\bibinfo {year}
  {2018})}\BibitemShut {NoStop}%
\bibitem [{\citenamefont {Tian}\ \emph {et~al.}(2021)\citenamefont {Tian},
  \citenamefont {Pan}, \citenamefont {Wang}, \citenamefont {Ye}, \citenamefont
  {Sheng}, \citenamefont {Wang}, \citenamefont {Juan}, \citenamefont {Huang},
  \citenamefont {Zhang}, \citenamefont {Xu} \emph {et~al.}}]{tian2021prb}%
  \BibitemOpen
  \bibfield  {author} {\bibinfo {author} {\bibfnamefont {C.~K.}\ \bibnamefont
  {Tian}}, \bibinfo {author} {\bibfnamefont {F.~H.}\ \bibnamefont {Pan}},
  \bibinfo {author} {\bibfnamefont {L.}~\bibnamefont {Wang}}, \bibinfo {author}
  {\bibfnamefont {D.~H.}\ \bibnamefont {Ye}}, \bibinfo {author} {\bibfnamefont
  {J.~M.}\ \bibnamefont {Sheng}}, \bibinfo {author} {\bibfnamefont {J.~C.}\
  \bibnamefont {Wang}}, \bibinfo {author} {\bibfnamefont {L.~J.}\ \bibnamefont
  {Juan}}, \bibinfo {author} {\bibfnamefont {J.~L.}\ \bibnamefont {Huang}},
  \bibinfo {author} {\bibfnamefont {H.~X.}\ \bibnamefont {Zhang}}, \bibinfo
  {author} {\bibfnamefont {D.~Y.}\ \bibnamefont {Xu}}, \emph {et~al.},\
  }\href@noop {} {\bibfield  {journal} {\bibinfo  {journal} {Phys. Rev. B}\
  }\textbf {\bibinfo {volume} {104}},\ \bibinfo {pages} {214410} (\bibinfo
  {year} {2021})}\BibitemShut {NoStop}%
\bibitem [{\citenamefont {Naguib}\ \emph {et~al.}(2011)\citenamefont {Naguib},
  \citenamefont {Kurtoglu}, \citenamefont {Presser}, \citenamefont {Lu},
  \citenamefont {Niu}, \citenamefont {Heon}, \citenamefont {Hultman},
  \citenamefont {Gogotsi},\ and\ \citenamefont {Barsoum}}]{naguib2011AM}%
  \BibitemOpen
  \bibfield  {author} {\bibinfo {author} {\bibfnamefont {M.}~\bibnamefont
  {Naguib}}, \bibinfo {author} {\bibfnamefont {M.}~\bibnamefont {Kurtoglu}},
  \bibinfo {author} {\bibfnamefont {V.}~\bibnamefont {Presser}}, \bibinfo
  {author} {\bibfnamefont {J.}~\bibnamefont {Lu}}, \bibinfo {author}
  {\bibfnamefont {J.}~\bibnamefont {Niu}}, \bibinfo {author} {\bibfnamefont
  {M.}~\bibnamefont {Heon}}, \bibinfo {author} {\bibfnamefont {L.}~\bibnamefont
  {Hultman}}, \bibinfo {author} {\bibfnamefont {Y.}~\bibnamefont {Gogotsi}},\
  and\ \bibinfo {author} {\bibfnamefont {M.~W.}\ \bibnamefont {Barsoum}},\
  }\href@noop {} {\bibfield  {journal} {\bibinfo  {journal} {Adv. Mater.}\
  }\textbf {\bibinfo {volume} {23}},\ \bibinfo {pages} {4248} (\bibinfo {year}
  {2011})}\BibitemShut {NoStop}%
\bibitem [{\citenamefont {Miranda}\ \emph {et~al.}(2016)\citenamefont
  {Miranda}, \citenamefont {Halim}, \citenamefont {Barsoum},\ and\
  \citenamefont {Lorke}}]{MA2016apl}%
  \BibitemOpen
  \bibfield  {author} {\bibinfo {author} {\bibfnamefont {A.}~\bibnamefont
  {Miranda}}, \bibinfo {author} {\bibfnamefont {J.}~\bibnamefont {Halim}},
  \bibinfo {author} {\bibfnamefont {M.~W.}\ \bibnamefont {Barsoum}},\ and\
  \bibinfo {author} {\bibfnamefont {A.}~\bibnamefont {Lorke}},\ }\href@noop {}
  {\bibfield  {journal} {\bibinfo  {journal} {Appl. Phys. Lett.}\ }\textbf
  {\bibinfo {volume} {108}},\ \bibinfo {pages} {033102} (\bibinfo {year}
  {2016})}\BibitemShut {NoStop}%
\bibitem [{\citenamefont {Kumar}\ \emph {et~al.}(2017)\citenamefont {Kumar},
  \citenamefont {Frey}, \citenamefont {Dong}, \citenamefont {Anasori},
  \citenamefont {Gogotsi},\ and\ \citenamefont {Shenoy}}]{kumar2017ACS}%
  \BibitemOpen
  \bibfield  {author} {\bibinfo {author} {\bibfnamefont {H.}~\bibnamefont
  {Kumar}}, \bibinfo {author} {\bibfnamefont {N.~C.}\ \bibnamefont {Frey}},
  \bibinfo {author} {\bibfnamefont {L.}~\bibnamefont {Dong}}, \bibinfo {author}
  {\bibfnamefont {B.}~\bibnamefont {Anasori}}, \bibinfo {author} {\bibfnamefont
  {Y.}~\bibnamefont {Gogotsi}},\ and\ \bibinfo {author} {\bibfnamefont {V.~B.}\
  \bibnamefont {Shenoy}},\ }\href@noop {} {\bibfield  {journal} {\bibinfo
  {journal} {ACS Nano}\ }\textbf {\bibinfo {volume} {11}},\ \bibinfo {pages}
  {7648} (\bibinfo {year} {2017})}\BibitemShut {NoStop}%
\bibitem [{\citenamefont {Wang}(2016)}]{wang2016JPCC}%
  \BibitemOpen
  \bibfield  {author} {\bibinfo {author} {\bibfnamefont {G.}~\bibnamefont
  {Wang}},\ }\href@noop {} {\bibfield  {journal} {\bibinfo  {journal} {J. Phys.
  Chem. C}\ }\textbf {\bibinfo {volume} {120}},\ \bibinfo {pages} {18850}
  (\bibinfo {year} {2016})}\BibitemShut {NoStop}%
\bibitem [{\citenamefont {Li}\ \emph {et~al.}(2021)\citenamefont {Li},
  \citenamefont {He}, \citenamefont {Grajciar},\ and\ \citenamefont
  {Nachtigall}}]{li2021JMCC}%
  \BibitemOpen
  \bibfield  {author} {\bibinfo {author} {\bibfnamefont {S.}~\bibnamefont
  {Li}}, \bibinfo {author} {\bibfnamefont {J.~J.}\ \bibnamefont {He}}, \bibinfo
  {author} {\bibfnamefont {L.}~\bibnamefont {Grajciar}},\ and\ \bibinfo
  {author} {\bibfnamefont {P.}~\bibnamefont {Nachtigall}},\ }\href@noop {}
  {\bibfield  {journal} {\bibinfo  {journal} {J. Mater. Chem. C}\ }\textbf
  {\bibinfo {volume} {9}},\ \bibinfo {pages} {11132} (\bibinfo {year}
  {2021})}\BibitemShut {NoStop}%
\bibitem [{\citenamefont {Weng}\ \emph {et~al.}(2016)\citenamefont {Weng},
  \citenamefont {Lin}, \citenamefont {Dagotto},\ and\ \citenamefont
  {Dong}}]{weng2016prl}%
  \BibitemOpen
  \bibfield  {author} {\bibinfo {author} {\bibfnamefont {Y.~K.}\ \bibnamefont
  {Weng}}, \bibinfo {author} {\bibfnamefont {L.~F.}\ \bibnamefont {Lin}},
  \bibinfo {author} {\bibfnamefont {E.}~\bibnamefont {Dagotto}},\ and\ \bibinfo
  {author} {\bibfnamefont {S.}~\bibnamefont {Dong}},\ }\href@noop {} {\bibfield
   {journal} {\bibinfo  {journal} {Phys. Rev. Lett.}\ }\textbf {\bibinfo
  {volume} {117}},\ \bibinfo {pages} {037601} (\bibinfo {year}
  {2016})}\BibitemShut {NoStop}%
\bibitem [{\citenamefont {Kresse}\ and\ \citenamefont
  {Furthm{\"u}ller}(1996)}]{kresse1996Prb}%
  \BibitemOpen
  \bibfield  {author} {\bibinfo {author} {\bibfnamefont {G.}~\bibnamefont
  {Kresse}}\ and\ \bibinfo {author} {\bibfnamefont {J.}~\bibnamefont
  {Furthm{\"u}ller}},\ }\href@noop {} {\bibfield  {journal} {\bibinfo
  {journal} {Phys. Rev. B}\ }\textbf {\bibinfo {volume} {54}},\ \bibinfo
  {pages} {11169} (\bibinfo {year} {1996})}\BibitemShut {NoStop}%
\bibitem [{\citenamefont {Perdew}\ \emph {et~al.}(1996)\citenamefont {Perdew},
  \citenamefont {Burke},\ and\ \citenamefont {Ernzerhof}}]{perdew1996Prl}%
  \BibitemOpen
  \bibfield  {author} {\bibinfo {author} {\bibfnamefont {J.~P.}\ \bibnamefont
  {Perdew}}, \bibinfo {author} {\bibfnamefont {K.}~\bibnamefont {Burke}},\ and\
  \bibinfo {author} {\bibfnamefont {M.}~\bibnamefont {Ernzerhof}},\ }\href@noop
  {} {\bibfield  {journal} {\bibinfo  {journal} {Phys. Rev. Lett.}\ }\textbf
  {\bibinfo {volume} {77}},\ \bibinfo {pages} {3865} (\bibinfo {year}
  {1996})}\BibitemShut {NoStop}%
\bibitem [{\citenamefont {Dudarev}\ \emph {et~al.}(1998)\citenamefont
  {Dudarev}, \citenamefont {Botton}, \citenamefont {Savrasov}, \citenamefont
  {Humphreys},\ and\ \citenamefont {Sutton}}]{dudarev1998Prb}%
  \BibitemOpen
  \bibfield  {author} {\bibinfo {author} {\bibfnamefont {S.~L.}\ \bibnamefont
  {Dudarev}}, \bibinfo {author} {\bibfnamefont {G.~A.}\ \bibnamefont {Botton}},
  \bibinfo {author} {\bibfnamefont {S.~Y.}\ \bibnamefont {Savrasov}}, \bibinfo
  {author} {\bibfnamefont {C.~J.}\ \bibnamefont {Humphreys}},\ and\ \bibinfo
  {author} {\bibfnamefont {A.~P.}\ \bibnamefont {Sutton}},\ }\href@noop {}
  {\bibfield  {journal} {\bibinfo  {journal} {Phys. Rev. B}\ }\textbf {\bibinfo
  {volume} {57}},\ \bibinfo {pages} {1505} (\bibinfo {year}
  {1998})}\BibitemShut {NoStop}%
\bibitem [{\citenamefont {He}\ \emph {et~al.}(2016)\citenamefont {He},
  \citenamefont {Lyu}, \citenamefont {Sun} \emph {et~al.}}]{he2016JMCC}%
  \BibitemOpen
  \bibfield  {author} {\bibinfo {author} {\bibfnamefont {J.}~\bibnamefont
  {He}}, \bibinfo {author} {\bibfnamefont {P.}~\bibnamefont {Lyu}}, \bibinfo
  {author} {\bibfnamefont {L.~Z.}\ \bibnamefont {Sun}}, \emph {et~al.},\
  }\href@noop {} {\bibfield  {journal} {\bibinfo  {journal} {J. Mater. Chem.
  C}\ }\textbf {\bibinfo {volume} {4}},\ \bibinfo {pages} {6500} (\bibinfo
  {year} {2016})}\BibitemShut {NoStop}%
\bibitem [{\citenamefont {Togo}\ and\ \citenamefont
  {Tanaka}(2015)}]{togo2015SM}%
  \BibitemOpen
  \bibfield  {author} {\bibinfo {author} {\bibfnamefont {A.}~\bibnamefont
  {Togo}}\ and\ \bibinfo {author} {\bibfnamefont {I.}~\bibnamefont {Tanaka}},\
  }\href@noop {} {\bibfield  {journal} {\bibinfo  {journal} {Scr. Mater.}\
  }\textbf {\bibinfo {volume} {108}},\ \bibinfo {pages} {1} (\bibinfo {year}
  {2015})}\BibitemShut {NoStop}%
\bibitem [{\citenamefont {Grimme}(2006)}]{grimme2006JCC}%
  \BibitemOpen
  \bibfield  {author} {\bibinfo {author} {\bibfnamefont {S.}~\bibnamefont
  {Grimme}},\ }\href@noop {} {\bibfield  {journal} {\bibinfo  {journal} {J.
  Comput. Chem.}\ }\textbf {\bibinfo {volume} {27}},\ \bibinfo {pages} {1787}
  (\bibinfo {year} {2006})}\BibitemShut {NoStop}%
\bibitem [{\citenamefont {Gulans}\ \emph {et~al.}(2014)\citenamefont {Gulans},
  \citenamefont {Kontur}, \citenamefont {Meisenbichler}, \citenamefont {Nabok},
  \citenamefont {Pavone}, \citenamefont {Rigamonti}, \citenamefont
  {Sagmeister}, \citenamefont {Werner},\ and\ \citenamefont
  {Draxl}}]{Gulans2014JPCM}%
  \BibitemOpen
  \bibfield  {author} {\bibinfo {author} {\bibfnamefont {A.}~\bibnamefont
  {Gulans}}, \bibinfo {author} {\bibfnamefont {S.}~\bibnamefont {Kontur}},
  \bibinfo {author} {\bibfnamefont {C.}~\bibnamefont {Meisenbichler}}, \bibinfo
  {author} {\bibfnamefont {D.}~\bibnamefont {Nabok}}, \bibinfo {author}
  {\bibfnamefont {P.}~\bibnamefont {Pavone}}, \bibinfo {author} {\bibfnamefont
  {S.}~\bibnamefont {Rigamonti}}, \bibinfo {author} {\bibfnamefont
  {S.}~\bibnamefont {Sagmeister}}, \bibinfo {author} {\bibfnamefont
  {U.}~\bibnamefont {Werner}},\ and\ \bibinfo {author} {\bibfnamefont
  {C.}~\bibnamefont {Draxl}},\ }\href@noop {} {\bibfield  {journal} {\bibinfo
  {journal} {J. Phys. Condens. Matter}\ }\textbf {\bibinfo {volume} {26}},\
  \bibinfo {pages} {363202} (\bibinfo {year} {2014})}\BibitemShut {NoStop}%
\bibitem [{\citenamefont {Sagmeister}\ and\ \citenamefont
  {Ambrosch-Draxl}(2009)}]{sagmeister2009PCCP}%
  \BibitemOpen
  \bibfield  {author} {\bibinfo {author} {\bibfnamefont {S.}~\bibnamefont
  {Sagmeister}}\ and\ \bibinfo {author} {\bibfnamefont {C.}~\bibnamefont
  {Ambrosch-Draxl}},\ }\href@noop {} {\bibfield  {journal} {\bibinfo  {journal}
  {Phys. Chem. Chem. Phys.}\ }\textbf {\bibinfo {volume} {11}},\ \bibinfo
  {pages} {4451} (\bibinfo {year} {2009})}\BibitemShut {NoStop}%
\bibitem [{\citenamefont {Landau}\ and\ \citenamefont
  {Binder}(2021)}]{landau2021guide}%
  \BibitemOpen
  \bibfield  {author} {\bibinfo {author} {\bibfnamefont {D.}~\bibnamefont
  {Landau}}\ and\ \bibinfo {author} {\bibfnamefont {K.}~\bibnamefont
  {Binder}},\ }\href@noop {} {\emph {\bibinfo {title} {A Guide to Monte Carlo
  Simulations in Statistical Physics}}}\ (\bibinfo  {publisher} {Cambridge
  university press},\ \bibinfo {year} {2021})\BibitemShut {NoStop}%
\bibitem [{\citenamefont {Khazaei}\ \emph {et~al.}(2013)\citenamefont
  {Khazaei}, \citenamefont {Arai}, \citenamefont {Sasaki}, \citenamefont
  {Chung}, \citenamefont {Venkataramanan}, \citenamefont {Estili},
  \citenamefont {Sakka},\ and\ \citenamefont {Kawazoe}}]{khazaei2013AFM}%
  \BibitemOpen
  \bibfield  {author} {\bibinfo {author} {\bibfnamefont {M.}~\bibnamefont
  {Khazaei}}, \bibinfo {author} {\bibfnamefont {M.}~\bibnamefont {Arai}},
  \bibinfo {author} {\bibfnamefont {T.}~\bibnamefont {Sasaki}}, \bibinfo
  {author} {\bibfnamefont {C.-Y.}\ \bibnamefont {Chung}}, \bibinfo {author}
  {\bibfnamefont {N.~S.}\ \bibnamefont {Venkataramanan}}, \bibinfo {author}
  {\bibfnamefont {M.}~\bibnamefont {Estili}}, \bibinfo {author} {\bibfnamefont
  {Y.}~\bibnamefont {Sakka}},\ and\ \bibinfo {author} {\bibfnamefont
  {Y.}~\bibnamefont {Kawazoe}},\ }\href@noop {} {\bibfield  {journal} {\bibinfo
   {journal} {Adv. Funct. Mater.}\ }\textbf {\bibinfo {volume} {23}},\ \bibinfo
  {pages} {2185} (\bibinfo {year} {2013})}\BibitemShut {NoStop}%
\bibitem [{\citenamefont {Zhao}\ \emph {et~al.}(2022)\citenamefont {Zhao},
  \citenamefont {Liu}, \citenamefont {Wang}, \citenamefont {Yang},
  \citenamefont {Bellaiche},\ and\ \citenamefont {Ma}}]{zhao2022prl}%
  \BibitemOpen
  \bibfield  {author} {\bibinfo {author} {\bibfnamefont {H.~J.}\ \bibnamefont
  {Zhao}}, \bibinfo {author} {\bibfnamefont {X.}~\bibnamefont {Liu}}, \bibinfo
  {author} {\bibfnamefont {Y.}~\bibnamefont {Wang}}, \bibinfo {author}
  {\bibfnamefont {Y.}~\bibnamefont {Yang}}, \bibinfo {author} {\bibfnamefont
  {L.}~\bibnamefont {Bellaiche}},\ and\ \bibinfo {author} {\bibfnamefont
  {Y.}~\bibnamefont {Ma}},\ }\href@noop {} {\bibfield  {journal} {\bibinfo
  {journal} {Phys. Rev. Lett.}\ }\textbf {\bibinfo {volume} {129}},\ \bibinfo
  {pages} {187602} (\bibinfo {year} {2022})}\BibitemShut {NoStop}%
\bibitem [{\citenamefont {Diwekar}\ \emph {et~al.}(2004)\citenamefont
  {Diwekar}, \citenamefont {Kamaev}, \citenamefont {Shi},\ and\ \citenamefont
  {Vardeny}}]{DM2004apl}%
  \BibitemOpen
  \bibfield  {author} {\bibinfo {author} {\bibfnamefont {M.}~\bibnamefont
  {Diwekar}}, \bibinfo {author} {\bibfnamefont {V.}~\bibnamefont {Kamaev}},
  \bibinfo {author} {\bibfnamefont {J.}~\bibnamefont {Shi}},\ and\ \bibinfo
  {author} {\bibfnamefont {Z.~V.}\ \bibnamefont {Vardeny}},\ }\href@noop {}
  {\bibfield  {journal} {\bibinfo  {journal} {Appl. Phys. Lett.}\ }\textbf
  {\bibinfo {volume} {84}},\ \bibinfo {pages} {3112} (\bibinfo {year}
  {2004})}\BibitemShut {NoStop}%
\bibitem [{\citenamefont {Yang}\ \emph {et~al.}(2020)\citenamefont {Yang},
  \citenamefont {Hu}, \citenamefont {Wu}, \citenamefont {Whangbo},
  \citenamefont {Radaelli},\ and\ \citenamefont {Stroppa}}]{Ke2020ACS}%
  \BibitemOpen
  \bibfield  {author} {\bibinfo {author} {\bibfnamefont {K.}~\bibnamefont
  {Yang}}, \bibinfo {author} {\bibfnamefont {W.~T.}\ \bibnamefont {Hu}},
  \bibinfo {author} {\bibfnamefont {H.}~\bibnamefont {Wu}}, \bibinfo {author}
  {\bibfnamefont {M.-H.}\ \bibnamefont {Whangbo}}, \bibinfo {author}
  {\bibfnamefont {P.~G.}\ \bibnamefont {Radaelli}},\ and\ \bibinfo {author}
  {\bibfnamefont {A.}~\bibnamefont {Stroppa}},\ }\href@noop {} {\bibfield
  {journal} {\bibinfo  {journal} {ACS Appl. Electron. Mater.}\ }\textbf
  {\bibinfo {volume} {2}},\ \bibinfo {pages} {1373} (\bibinfo {year}
  {2020})}\BibitemShut {NoStop}%
\bibitem [{\citenamefont {Sangalli}\ \emph {et~al.}(2012)\citenamefont
  {Sangalli}, \citenamefont {Marini},\ and\ \citenamefont
  {Debernardi}}]{sangalli2012prb}%
  \BibitemOpen
  \bibfield  {author} {\bibinfo {author} {\bibfnamefont {D.}~\bibnamefont
  {Sangalli}}, \bibinfo {author} {\bibfnamefont {A.}~\bibnamefont {Marini}},\
  and\ \bibinfo {author} {\bibfnamefont {A.}~\bibnamefont {Debernardi}},\
  }\href@noop {} {\bibfield  {journal} {\bibinfo  {journal} {Phys. Rev. B}\
  }\textbf {\bibinfo {volume} {86}},\ \bibinfo {pages} {125139} (\bibinfo
  {year} {2012})}\BibitemShut {NoStop}%
\bibitem [{\citenamefont {Born}\ and\ \citenamefont
  {Wolf}(2013)}]{born2013principles}%
  \BibitemOpen
  \bibfield  {author} {\bibinfo {author} {\bibfnamefont {M.}~\bibnamefont
  {Born}}\ and\ \bibinfo {author} {\bibfnamefont {E.}~\bibnamefont {Wolf}},\
  }\href@noop {} {\emph {\bibinfo {title} {Principles of optics:
  electromagnetic theory of propagation, interference and diffraction of
  light}}}\ (\bibinfo  {publisher} {Elsevier},\ \bibinfo {year}
  {2013})\BibitemShut {NoStop}%
\bibitem [{\citenamefont {Chandrasekaran}\ \emph {et~al.}(2017)\citenamefont
  {Chandrasekaran}, \citenamefont {Mishra},\ and\ \citenamefont
  {Singh}}]{Chandrasekaran2017nl}%
  \BibitemOpen
  \bibfield  {author} {\bibinfo {author} {\bibfnamefont {A.}~\bibnamefont
  {Chandrasekaran}}, \bibinfo {author} {\bibfnamefont {A.}~\bibnamefont
  {Mishra}},\ and\ \bibinfo {author} {\bibfnamefont {A.~K.}\ \bibnamefont
  {Singh}},\ }\href@noop {} {\bibfield  {journal} {\bibinfo  {journal} {Nano
  Lett.}\ }\textbf {\bibinfo {volume} {17}},\ \bibinfo {pages} {3290} (\bibinfo
  {year} {2017})}\BibitemShut {NoStop}%
\bibitem [{\citenamefont {An}\ \emph {et~al.}(2019)\citenamefont {An},
  \citenamefont {Yand}, \citenamefont {Jun}, \citenamefont {Min}, \citenamefont
  {Jun},\ and\ \citenamefont {Dong}}]{An2019JPCC}%
  \BibitemOpen
  \bibfield  {author} {\bibinfo {author} {\bibfnamefont {M.}~\bibnamefont
  {An}}, \bibinfo {author} {\bibfnamefont {Z.}~\bibnamefont {Yand}}, \bibinfo
  {author} {\bibfnamefont {C.}~\bibnamefont {Jun}}, \bibinfo {author}
  {\bibfnamefont {Z.~H.}\ \bibnamefont {Min}}, \bibinfo {author} {\bibfnamefont
  {G.~Y.}\ \bibnamefont {Jun}},\ and\ \bibinfo {author} {\bibfnamefont
  {S.}~\bibnamefont {Dong}},\ }\href@noop {} {\bibfield  {journal} {\bibinfo
  {journal} {J. Phys. Chem. C}\ }\textbf {\bibinfo {volume} {123}},\ \bibinfo
  {pages} {30545} (\bibinfo {year} {2019})}\BibitemShut {NoStop}%
\bibitem [{\citenamefont {Kato}\ \emph {et~al.}(2004)\citenamefont {Kato},
  \citenamefont {Myers}, \citenamefont {Gossard},\ and\ \citenamefont
  {Awschalom}}]{kato2004SCI}%
  \BibitemOpen
  \bibfield  {author} {\bibinfo {author} {\bibfnamefont {Y.~K.}\ \bibnamefont
  {Kato}}, \bibinfo {author} {\bibfnamefont {R.~C.}\ \bibnamefont {Myers}},
  \bibinfo {author} {\bibfnamefont {A.~C.}\ \bibnamefont {Gossard}},\ and\
  \bibinfo {author} {\bibfnamefont {D.~D.}\ \bibnamefont {Awschalom}},\
  }\href@noop {} {\bibfield  {journal} {\bibinfo  {journal} {Science}\ }\textbf
  {\bibinfo {volume} {306}},\ \bibinfo {pages} {1910} (\bibinfo {year}
  {2004})}\BibitemShut {NoStop}%
\bibitem [{\citenamefont {Lee}\ \emph {et~al.}(2016)\citenamefont {Lee},
  \citenamefont {Mak},\ and\ \citenamefont {Shan}}]{Lee2016NN}%
  \BibitemOpen
  \bibfield  {author} {\bibinfo {author} {\bibfnamefont {J.}~\bibnamefont
  {Lee}}, \bibinfo {author} {\bibfnamefont {K.~F.}\ \bibnamefont {Mak}},\ and\
  \bibinfo {author} {\bibfnamefont {J.}~\bibnamefont {Shan}},\ }\href@noop {}
  {\bibfield  {journal} {\bibinfo  {journal} {Nat. Nanotechnol.}\ }\textbf
  {\bibinfo {volume} {11}},\ \bibinfo {pages} {421} (\bibinfo {year}
  {2016})}\BibitemShut {NoStop}%
\end{thebibliography}%
\bibliographystyle{apsrev4-2}
\end{document}